\documentclass{IEEEtaes}

\usepackage{color,array,amsthm}
\usepackage{graphicx}
\usepackage{amssymb}
\usepackage{amsmath}
\usepackage{algorithm}
\usepackage{subcaption}
\usepackage{algpseudocode}
\usepackage{makecell}
\usepackage{multirow}

\doiinfo{TAES.2022.Doi Number}

\usepackage{bm}

\begin{document}

\title{Satellites swarm cooperation for pursuit-attachment tasks with transformer-based reinforcement learning}

\author{YONGHAO LI}
\affil{School of Automation, Beijing University of Posts and Telecommunications, Beijing 100876, China}

%% \author{FOURTH D. AUTHOR}
%%\member{IEEE}
%% \affil{University of Colorado, Colorado, USA}

%% \accepteddate{XXXXX XX XXXX}
%% \publisheddate{XXXXX XX XXXX}

\maketitle

\begin{abstract}
The on-orbit intelligent planning of satellites swarm has attracted increasing attention from scholars. Especially in tasks such as the pursuit and attachment of non-cooperative satellites, satellites swarm must achieve coordinated cooperation with limited resources. 
The study proposes a reinforcement learning framework that integrates the transformer and expert networks.
Firstly, under the constraints of incomplete information about non-cooperative satellites, an implicit multi-satellites cooperation strategy was designed using a communication sharing mechanism.
Subsequently, for the characteristics of the pursuit-attachment tasks, 
the multi-agent reinforcement learning framework is improved by introducing transformers and expert networks inspired by transfer learning ideas.
To address the issue of satellites swarm scalability, sequence modelling based on transformers is utilized to craft memory-augmented policy networks, meanwhile   increasing the scalability of the swarm.
By comparing the convergence curves with other algorithms, it is shown that the proposed method is qualified for pursuit-attachment tasks of satellites swarm.
Additionally, simulations under different maneuvering strategies of non-cooperative satellites respectively demonstrate the robustness of the algorithm and the task efficiency of the swarm system. The success rate of pursuit-attachment tasks is analyzed through Monte Carlo simulations.
\end{abstract}

\section{INTRODUCTION}

In recent years, satellites swarm has gained widespread attention due to its increasing significance in areas such as space dominance and information competition\cite{bib1}. Meanwhile, intelligent planning for satellite rendezvous and proximity operations has been an active research area and shows great potential in satellite refueling, formation flying, and orbital debris mitigation and avoidance\cite{bib2}. Typically, the satellites swarm system is composed of multiple satellites that work together to accomplish tasks, such as pursuing and attaching to a satellite.However, Several challenges  affect the performances of satellites swarm for the tasks. The execution of pursuit-attachment tasks has become increasingly challenging due to the enhancing capabilities of non-cooperative satellites autonomous decision-making and evasion. In order to alleviate the workload of ground systems, satellites swarm should be equipped with multitasking autonomous cooperation algorithms and real-time intelligent decision-making capabilities\cite{bib3}.

The pursuit-attachment tasks mentioned above encounter various challenges, i.e. short task cycles, multiple real-time constraints, and complex constraint propagation.
It is tough to obtain knowledge about the maneuvering strategies and intentions of non-cooperative satellites, the satellites swarm is unable to predict multiple time-varying trajectories through inertial measurements, especially during high-speed movements. Due to sensor perception limitations, on-board sensors do not receive full states from nearby satellites, this limits our grasp of global information. These make the application of swarm intelligent cooperation crucial.In study\cite{bib4}, the authors proposed nonlinear control laws to address the pursuit-evasion game problem, leveraging state-dependent Riccati methods and linear quadratic differential game theory. Ye et al.\cite{bib6} proposed that under different configurations for both pursuing and evading, high computational complexity is encountered when seeking open-loop solutions to nonlinear equations through iterative searching. Gong et al. \cite{bib5} introduced a reachable domain to assist in problem analysis. Kartal et al. \cite{bib8} proposed integral RL under velocity constraints to achieve a satellite rendezvous with a target satellite.
Zong et al.\cite{bib32} used the calculus of variational method to solve the optimal control problem to achieve satellite rendezvous and capturing satellites. Wu et al. \cite{bib33} utilized multiple microsatellites to implement takeover control of spacecraft using differential game methods. The attachment task is a prerequisite for the subsequent rendezvous, capturing, and control takeover.

Additionally, satellites swarm pursuit-attachment two-stage tasks under resource constraints have temporal dependencies. Traditional Markov decision paradigms overlook long-short term historical temporal information, leading to Decentralized partially observable Markov decision processes (Dec-POMDPs) problems. Meanwhile, as the number of intelligent agents increases, the above solutions may be risks of performance degradation and system crash, making it difficult for us to extend them to larger-scale swarm tasks, and strategy switching between tasks of different scales.
Addressing non-cooperative satellites and multi-task problems under incomplete information. Model-based control methods are difficult to complete the capture-attachment multi-task \cite{bib9,bib23,bib33}. Many scholars have adopted reinforcement learning (RL) methods for research \cite{bib11,bib25,bib26,bib28,bib34}. However, these methods are not suitable for long-term planning problems, and the importance of scalability and memorability of swarm algorithms in performance improvement has been exposed. Among them, papers \cite{bib26,bib28} focused on the task optimization under resource constraints. Paper \cite{bib34} uses attention modules to generate the relationship between tasks and expert guidance networks. Table \ref{tab1} summarizes detailed comparisons of the literatures discussed above.

\begin{table*}[!h]
	\centering
	\caption{Comparison of relevant literatures}
 	\label{tab1}
	\begin{tabular*}{0.95\textwidth}{@{\extracolsep{\fill}}cccccc@{}}
		\hline
		\textbf{Refs.} & \textbf{Research Topic} & \textbf{Method} &\makecell{\textbf{Pursuers}\\ \textbf{and evaders}} & \makecell{\textbf{Cooperation} \\\textbf{(yes or not)}} & \textbf{Task type} \\
		\hline
	\cite{bib2}	& satellite pursuit-evasion & reachable analysis & \multirow{2}{*}{1 vs 1} & \multirow{6}{*}{N} & \multirow{10}{*}{pursuit-evasion} \\ \cline{1-3}
	\cite{bib8}	& satellite pursuit-evasion & IRL & &  & \\ \cline{1-4}
 	\cite{bib23}	& \makecell{multi-satellite\\ pursuit-evasion} & CNP & \multirow{7}{*}{n vs 1} &  & \\\cline{1-3}
	\cite{bib11}	& \makecell{UAVs multi-agent \\pursuit-evasion} & \multirow{3}{*}{MADDPG} &  &  &  \\ \cline{1-2}\cline{5-5}
	\cite{bib25}	& \makecell{multi-agent cooperation in\\ satellite pursuit-evasion} &  &  & \multirow{3}{*}{Y} &  \\\cline{1-3}
	\cite{bib26,bib28}	& \makecell{UAVs in multi-agent \\pursuit-evasion} & MAPPO &  &  &  \\\hline
	\cite{bib32}	& \makecell{satellite rendezvous\\ and capturing} & \makecell{calculus of \\variational method} & \multirow{3}{*}{/} & \multirow{3}{*}{/} & \multirow{3}{*}{attachment} \\ \cline{1-3}
	\cite{bib33}	& \makecell{multi-satellite\\ takeover control} & \makecell{differential\\ game methods} &  &  &  \\ \hline
		this study & \makecell{multi-satellite \\pursuit-evasion \\and attachment} & \makecell{MAPPO-expert\\ mixture transformer }& n vs m & Y & pursuit-attachment \\
		\hline
	\end{tabular*}
\end{table*}

 With the improvement of the autonomous maneuvering capability of non-cooperative satellites, the traditional convex optimization methods lack advantages in complexity and computational efficiency, rendering the effective accomplishment of the pursuit-attachment task difficult.
While the introduction of Deep Reinforcement Learning (DRL) fully utilizes the perceptual fitting capability of deep networks and the real-time decision-making ability of reinforcement learning, meanwhile also avoiding complex algorithm designs.
Deep neural networks(DNN) as a high-performance function approximation method have made significant progress in various fields such as agent vision, multi-agent systems, and context memory\cite{bib29,bib30,bib31}. And on the basis of the analysis and modeling results based on DNN, RL can allow more in-depth processes for complex reasoning, decision-making, control, long-term planning, and collaborative tasks involving multiple agents. The RL algorithm utilizes the interactions between the agent and environment to improve autonomous decision-making, which provides a novel approach to spatial swarm intelligence. Many scholars engage in related research. The papers\cite{bib10,bib11} proposed MADDPG and MAPPO algorithms to control swarm of unmanned aerial vehicles. These algorithms can handle dynamic environments and avoid obstacles. However, the utilized networks lack memory, resulting in poor performance in long-term planning. However, the aforementioned solutions exhibit subpar performance in long-term planning scenarios due to the absence of long-term memory in the neural networks utilized.
Therefore, autonomous decision based on DRL plays a crucial role in the work of pursuit-attachment tasks \cite{bib3}.

In recent years, Transformer, as a new architecture for DNN, has shown outstanding performance in a wide range of fields such as NLP, CV, etc. \cite{bib13,bib14}, 
excelling in handling long-distance dependencies \cite{bib15}, and benefiting from the advantages of parallel computation \cite{bib16}.
In paper\cite{bib12}, Dai et al. proposed a new architecture called GTrXL (Gated Transformer-XL). GTrXL can understand time dependencies that extend beyond a fixed length while preserving temporal continuity. This allows it to integrate current input trajectories with past trajectories for prediction. Janner et al. trained a self-supervised regression model on offline data in a purely supervised manner\cite{bib18}. The (Online) Decision Transformers \cite{bib19,bib20} generates future actions based on expected returns and past states and actions, which evades bypassing the complex dynamic programming calculations for cumulative rewards, bypassing dynamic programming calculations for cumulative rewards. 
The Transformer architecture has several notable advantages, including the ability to model long dependencies and excellent scalabilities\cite{bib17}. Additionally, it achieves strong performance in small-sample generalization tasks.

While these algorithms have been successful in their specific domains, there is limited research for satellites swarm based on RL methods. They may not be suitable for a satellites swarm multitasking scenario. Furthermore, conventional decision-making in traditional RL agents based on Markov decision processes, which only considers the current state, may neglect important temporal information and result in the Dec-POMDPs problem. In addition, it confronts challenges including low sample efficiency, vast exploration space, and inadequate behavioral generalization ability, ultimately leading to diminished task efficiency and success rates. Therefore, to address the Dec-POMDPs problem arising from neglecting temporal information, we propose the Multi-Attention architecture.

In this approach, expert networks are used for guiding based on the fine-grained characteristics of multiple tasks, and multi-agent technique is integrated. 
Autonomous cooperation on satellites swarm is carried out by sharing perception information and memorizing motion trajectories. And a close collaborative mode is facilitated. The main contributions are summarized as follows:

1) A  MAPPO-Expert Mixture Transformer (MAPPO-EMT) framework is proposed to deal with collaborative tasks on satellites swarm. To the best of our knowledge, this is the first time that integrates swarm perception computation and decision-making with a combination of Transformer sequence modeling and expert-guided networks. The framework initially enhances perceptual computing capabilities. Subsequently, it utilizes encoder networks to share effective temporal perceptual data, enabling decoder networks to quickly make optimal decisions in response to changes in objectives. It makes it possible for the main network to adaptively select for various tasks. They adaptively select appropriate expert modes for various tasks through the main network.

2) The model for satellites swarm system is constructed based on the Clohessy-Wiltshire(CW) model of orbital motion, which is characterized a set of ordinary differential equations (ODEs). And then the MAPPO-EMT can be employed deeply for the control of orbital dynamics decisions. Adopting this approach allows for fully leveraging the fitting potential of transformer for partial differential equations, as well as the long-short term memory capability and online learning, and real-time decision-making abilities of reinforcement learning.

3) Through simulation experiments, the proposed MAPPO-EMT outperforms traditional MADDPG and MAPPO in terms of convergence speed and average reward. The algorithm effectively addresses scalability issues arising from changes in the number of entities in three task scenarios with different settings.
Additionally, the algorithm's effectiveness is validated in the presence of multiple non-cooperative satellites.

The remaining part of this article is organized as follows. Section \ref{sec2} introduces the scene and model description; Section \ref{sec3} proposes a swarm collaborative decision-making framework based on sequential modeling of mixed expert knowledge; numerical simulations and analysis are carried out in Section \ref{sec4}; Section \ref{sec5} summarizes the entire article and provides prospects.

\section{SYSTEM MODELS}\label{sec2}

\begin{figure}[!h]
	\centerline{\includegraphics[width=19.5pc]{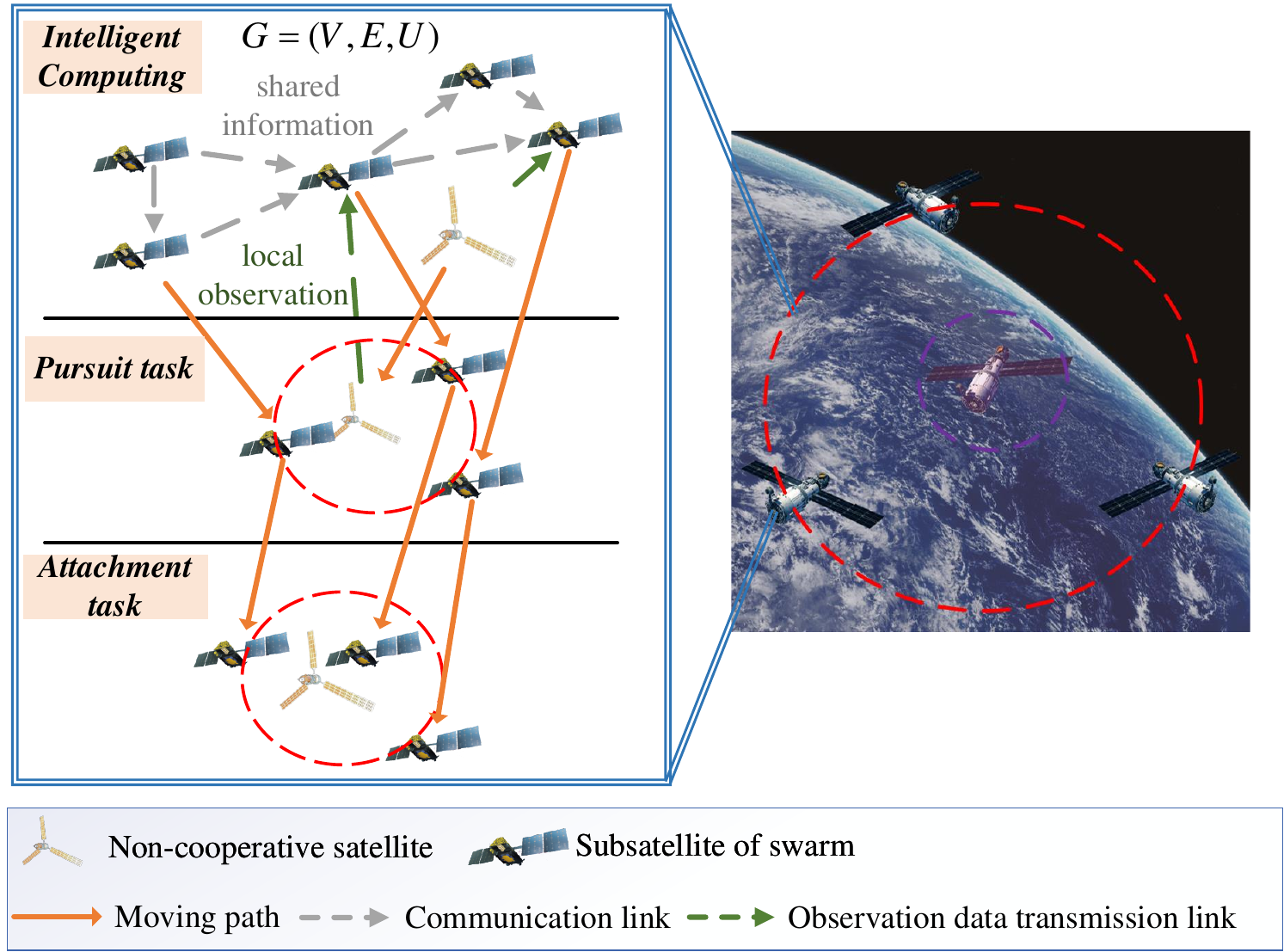}}
	\caption{Illustration of a satellite application scenario with swarm-based space cooperation tasks}\label{fig1}
\end{figure}

\begin{figure}[!h]
	\centerline{\includegraphics[width=19.5pc]{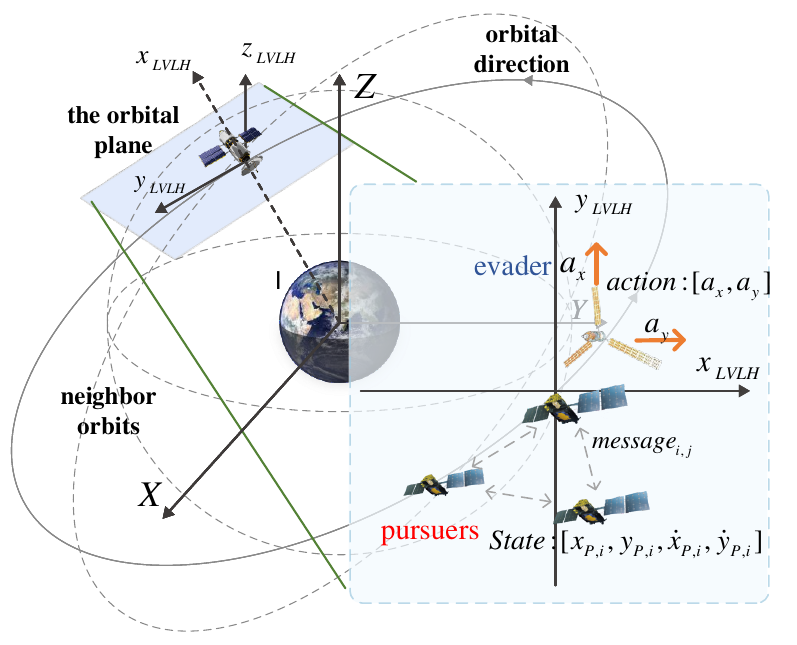}}
	\caption{Satellites swarm based on orbital kinematic model}\label{fig2}
\end{figure}

As shown in Figure \ref{fig1}, the scenarios of pursuit-attachment tasks are carried out, satellites swarm need to complete the task under the condition of incomplete information about the non-cooperative satellites' strategy. This can be described as an ordered and collision-free swarm intelligent task system on the  satellites swarm within a limited time. Firstly, the orbital kinematics model is established based on the CW relative motion equation. 
The architecture, as shown in Figure \ref{fig2}, considers a scenario with $k$=1 and $m$=3, where the non-cooperative satellites set is represented as $\bm{\mathcal{K}}=\{1,2, \ldots, k\}$. One can assume a swarm of $m$  satellites, denoted as $\bm{\mathcal{M}}=\{1,2, \ldots, m\}$, and their communication collaborations form a topology graph represented by $G=(V, E, U)$, where $V$ represents node satellites, $E$ represents satellite communication collaboration relationships, and $U$ represents global information such as the number of nodes.

\subsection{Orbital dynamics model}

The CW  equations are widely used in tasks such as designing algorithm for spacecraft close-range rendezvous guidance and control, as well as spacecraft formation flight configuration design and control \cite{bib21}. In order to describe the orbital kinematics of pursuit-attachment model, it is necessary to establish an orbital coordinate system on the reference orbit. Figure \ref{fig2} shows the local vertical local-horizontal (LVLH) coordinate system $\mathcal{L}_{x y z}$ \cite{bib8}. When the satellite is in close proximity to the reference satellite, the relative motion of satellite $i$ can be described using the CW equation:

\begin{equation}\label{eq1}
\left\{\begin{array}{l}
	\ddot{x}_i-2 \omega \dot{y}_i-3 \omega^{2} x_i=a_{ix} \\
	\ddot{y}_i+2 \omega \dot{x}_i=a_{iy} \\
	\ddot{z}_i+\omega^{2} z_i=a_{iz}
\end{array}\right.
\end{equation}

Where $\left[x_i, y_i, z_i\right]^T$ and $\left[\dot{x}_i, \dot{y}_i, \dot{z}_i\right]^T$ represent the three components of the satellite i  position and velocity  in the LVLH coordinate system. $a_{ix}$, $a_{iy}$, and $a_{iz}$ denote the three-axis thrust acceleration of the spacecraft along the $O_{x}, O_{y}, O_{z}$. $\omega \triangleq \sqrt{\mu / a_{0}^{3}}$ signifies the orbital angular velocity of the reference satellite, $\mu$ is the gravitational constant, and $a_0$ is the orbital radius, respectively. When in a geostationary satellite orbit, the value of $a_0$  can be 42164km. In order to better describe the design of the agent, one can choose state vector $\boldsymbol{X_i}=[x_i, y_i, z_i, \dot{x}_i, \dot{y}_i, \dot{z}_i]^{T}$ to describe the position and velocity in the LVLH coordinate system. The control vector, which represents the three control inputs, is given by $\boldsymbol{a_i}=\left[a_{ix}, a_{iy}, a_{iz}\right]^{T}$.

For the $i$-th satellite in the swarm, let state vector $\boldsymbol{X}_{i}(t)=\left[x_{i}(t), \dot{x}_{i}(t), y_{i}(t), \dot{y}_{i}(t), z_{i}(t), \dot{z}_{i}(t)\right]^{T}$, and the control vector can be written as $\boldsymbol{a}_{i}(t)=\left[a_{i, X}(t), a_{i, Y}(t), a_{i, Z}(t)\right]^{T}$.

Equation (\ref{eq1}) can be rewritten as 
\begin{equation}\label{eq2}
\dot{\boldsymbol{X}}_{i}(t)=\widetilde{\boldsymbol{A}} \boldsymbol{X}_{i}(t)+\widetilde{\boldsymbol{B}} \boldsymbol{a}_{i}(t)
\end{equation}
The matrices $\widetilde{\boldsymbol{A}}$ and $\widetilde{\boldsymbol{B}}$ can be found in Appendix.

For the pursuit-evasion tasks, when the pursuer  approaches the position of the evader , they can choose a virtual satellite near this close range as a reference satellite to establish the LVLH framework. Therefore, the relative kinematics can be expressed as follows:
\begin{equation}\label{eq3}
	\dot{\boldsymbol{X}}_{P, i}(t)  =\widetilde{\boldsymbol{A}} \boldsymbol{X}_{P, i}(t)+\widetilde{\boldsymbol{B}} \boldsymbol{a}_{P, i}(t) 
\end{equation}
\begin{equation}\label{eq4}
\dot{\boldsymbol{X}}_{E}(t) =\widetilde{\boldsymbol{A}} \boldsymbol{X}_{E}(t)+\widetilde{\boldsymbol{B}} \boldsymbol{a}_{E}(t)
\end{equation}
where $\boldsymbol{a}_{P, i}$ and $\boldsymbol{a}_{E}$ denote the control accelerations of the pursuer and evader satellites, respectively.

\subsection{Task queuing model}

During the pursuit-attachment task, the non-cooperative satellites are detected by  satellites swarm using onboard sensors and communication modules. 
These devices can rapidly collect multi-modal data and transmit it to the control processing unit for thorough analysis. Data processing, which refers to computing tasks, involves taking into account environmental information, the current state and position of the target, and the states of nearby  satellites. However, individual satellite has limited to compute capabilities, and intensive computing loads may cause high latency issues. In contrast, some  satellites may quickly derive computing results due to their narrow sensing range. To optimize the computation time of the  satellites swarm, we design a collaborative computing scheme.

The computational load of  satellite $i$ in timeslot $t$ can be denoted as $Q_{i}(t)$, $t \in\{1,2, \ldots, T\}$, and it consists of three parts:

1) Sensing data $S_i (t)$ collected by the onboard sensor of the  satellite $i$ in the current timeslot.

2) Data $y_{(i,j)} (t)$ transmitted by the  satellite $j$ for collaborative computation.

3) Residual data $Q_{i}(t-1)-y_{i}(t-1)$ from the previous timeslot, where $y_{i}(t-1)$ denoted as the computational load of  satellite $i$ at timeslot $t-1$.

\begin{figure}[!h]
	\centerline{\includegraphics[width=19.5pc]{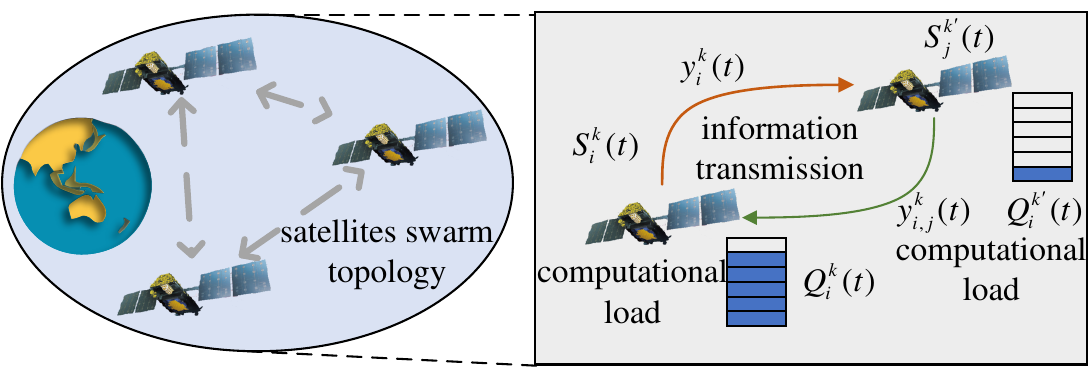}}
	\caption{Collaborative computing offloading diagram}\label{fig3}
\end{figure}

As shown in Fig. \ref{fig3}, all sensing data are represented as numerical values in timeslot $t$. the satellite $i$ perceives the computational task load $S_{i}^{k}(t)$ generated by target $k$. The task load $x_{i}^{k}(t)$ can be computed by satellite $i$ itself, while the task load $y_{i, j}^{k}(t)$ can be offloaded to nearby resource-available sub-satellite $j$.

After computing by satellite $j$, the results are transmitted back to satellite $i$ to accurately estimate the target state. Through this collaborative computing mechanism, satellite $i$ can complete the computation efficiently.

After satellite $i$ obtains sensing data from target $k$, it updates the queue size, which can be expressed as:
\begin{equation}\label{eq5}
Q_{i}^{k}(t+1)=\left[Q_{i}^{k}(t)-\sum_{j=1}^{M} a_{i, j} y_{i, j}^{k}(t)\right]^{+}+S_{i}^{k}(t)
\end{equation}
where, $a_{i,j}$ represents the collaborative computation offload from satellite $i$ to satellite $j$, with a value of 0 or 1, indicating the presence or absence of collaborative computation offload.

\section{PROBLEM FORMULATION AND APPROACH}\label{sec3}

This section introduces a multi-task reinforcement learning sequential modeling framework that integrates expert networks to specifically combine different expert knowledge for the pursuit and attachment substage tasks. The method can be described in two main parts: 1) an Encoder-Decoder backbone network based on the Attention mechanism for extracting temporal information; and 2) two expert networks for extracting task-specific pursuit-attachment representations.

\begin{figure*}[!h]
	\centerline{\includegraphics[width=39pc]{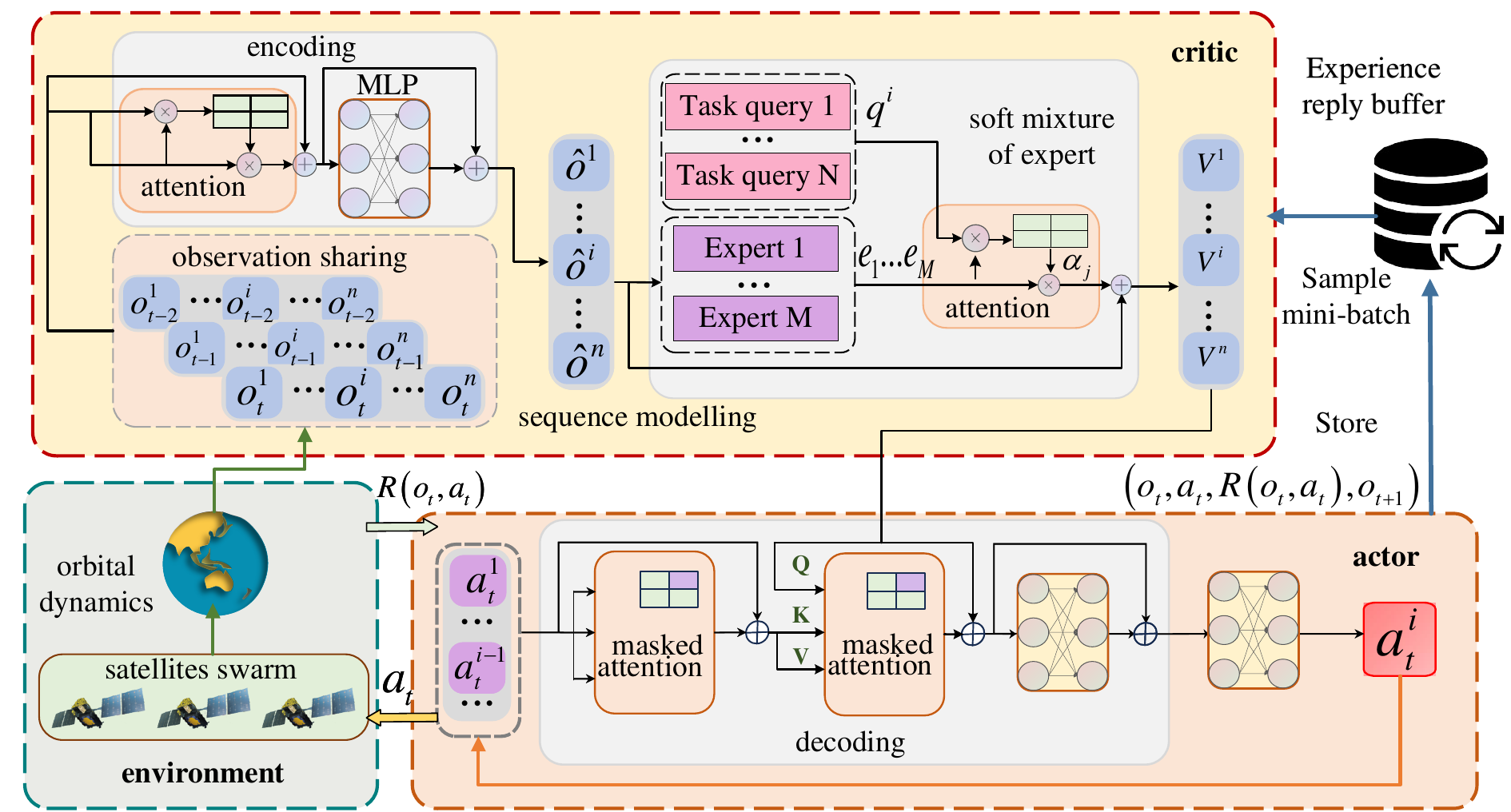}}
	\caption{Satellites swarm collaboration based on expert mixture transformer}\label{fig4}
\end{figure*}

\subsection{MARL sequences model}
To obtain the optimal control strategy for the agents, one can describe the system by using  Dec-POMDPs. We define a six-tuple: ${\langle\mathcal{N}, S,\{\boldsymbol{\mathcal { O }}\},\{\boldsymbol{\mathcal { A }}\}, R, P, \gamma\rangle}$ where ${\mathcal{N}=\{1, \ldots, {n}\}}$ represents the set of agents; ${S}$ is the set of states; ${\boldsymbol{\mathcal { O }}=\prod_{t=1}^{T} \prod_{i=1}^{n} \mathcal{O}_{t}^{i}}$ denotes the shared local observation space for the agents; ${\boldsymbol{\mathcal { A }}=\prod_{i=1}^{n} \mathcal{A}^{i}}$ represents the product of individual action spaces, i.e., the joint action space; ${R: \boldsymbol{\mathcal { O }} \times \boldsymbol{\mathcal { A }} \rightarrow\left[-R_{\max }, R_{\max }\right]}$ represents the joint reward function; ${P: \mathcal{O} \times \mathcal{A} \times \mathcal{O} \rightarrow \mathbb{R}}$ represents the state transition probability function, and ${\gamma \in[0,1)}$ is the discount factor. Specifically, due to the Dec-POMDP problem caused by ignoring temporal information and the inability to obtain the true state ${S}$, we base the observation space of the agents on sequence modeling. In detail, at time step ${t \in \mathbb{N}}$, sub-satellite ${i \in \mathcal{N}}$ observes ${o_{t}^{i} \in \mathcal{O}^{i}}$, and then ${\boldsymbol{o}_{\boldsymbol{t}}=\left({o}_{t}^{1}, \ldots, {o}_{t}^{i}, \ldots, {o}_{t}^{n}\right)}$ represents the joint observation. The input ${\left\{\boldsymbol{o}_{\boldsymbol{t}-\mathbf{2}}, \boldsymbol{o}_{\boldsymbol{t}-\mathbf{1}}, \boldsymbol{o}_{\boldsymbol{t}}\right\}}$ guides the selection of action ${a_{t}^{i}}$ based on the policy ${\pi^{i}}$, where all agents ${\mathcal{N}=\{1, \ldots, {n}\}}$ jointly contribute to the collective policy $\boldsymbol{\pi}$.

Agents evaluate the values of actions and observations through ${Q_{\pi}(\boldsymbol{o}, \boldsymbol{a})}$ and the value network ${V_{\boldsymbol{\pi}}(\boldsymbol{o})}$, defined as follows:
\begin{equation}\label{eq6}
\begin{array}{l}
	Q_{\boldsymbol{\pi}}(\boldsymbol{o}, \boldsymbol{a}) \triangleq \mathbb{E}_{\mathbf{o}_{1: \infty} \sim P, \boldsymbol{a}_{1: \infty} \sim \boldsymbol{\pi}}\left[R^{\gamma} \mid \mathbf{o}_{0}\!=\!\boldsymbol{o}, \boldsymbol{a}_{0}\!=\!\boldsymbol{a}\right] \\
	V_{\boldsymbol{\pi}}(\boldsymbol{o}) \triangleq \mathbb{E}_{\boldsymbol{a}_{0} \sim \boldsymbol{\pi}, \mathbf{o}_{1: \infty} \sim P, \boldsymbol{a}_{1: \infty} \sim \boldsymbol{\pi}}\left[R^{\gamma} \mid \mathbf{o}_{0}\!=\!\boldsymbol{o}\right]
\end{array}
\end{equation}

For any arbitrary, non-overlapping, ordered subsets of intelligent agents, $i_{1: m}=\left\{i_{1}, \ldots, i_{m}\right\}$ and $j_{1: h}=\left\{j_{1}, \ldots, j_{h}\right\}$, where $m, h \leq n$. The multi-agent observation-value functions can be defined as:
\begin{equation}\label{eq7}
Q_{\pi}\left(\boldsymbol{o}, \boldsymbol{a}^{i_{1: m}}\right) \triangleq \mathbb{E}\left[R^{\gamma} \mid \mathbf{o}_{0}^{i_{1: n}}=\boldsymbol{o}, \mathbf{a}_{0}^{i_{1: m}}=\boldsymbol{a}^{i_{1: m}}\right]
\end{equation}

Subsequently, to further quantify the contribution of the selected subset of intelligent agents to the joint rewards, we define the multi-agent advantage function as follows:
\begin{equation}\label{eq8}
\begin{aligned}
&A_{\boldsymbol{\pi}}^{i_{1: m}}\left(\boldsymbol{o}, \boldsymbol{a}^{j_{1: h}}, \boldsymbol{a}^{i_{1: m}}\right)\\
 \triangleq& Q_{\boldsymbol{\pi}}^{j_{1: h}, i_{1: m}}\left(\boldsymbol{o}, \boldsymbol{a}^{j_{1: h}}, \boldsymbol{a}^{i_{1: m}}\right)-Q_{\boldsymbol{\pi}}^{j_{1: h}}\left(\boldsymbol{o}, \boldsymbol{a}^{j_{1: h}}\right)
\end{aligned}
\end{equation}

The meaning of this is that intelligent agents ${i_{1: m}}$ taking joint action ${\boldsymbol{a}^{i_{1: m}}}$ is better or worse than intelligent agents ${i_{1: {h}}}$ taking action ${\boldsymbol{a}^{i_{1: h}}}$ by a certain amount. This value function allows for the study of interactions between them and decomposes the joint value function. By arranging intelligent agents ${i_{1: n}}$ and considering any joint observation ${\boldsymbol{o}=\boldsymbol{o} \in \boldsymbol{O}}$ and joint action ${\boldsymbol{a}=\boldsymbol{a}^{i_{1: n}} \in \bm{\mathcal{A}}}$, the multi-agent advantage decomposition theorem can be derived:
\begin{equation}\label{eq9}
A_{\boldsymbol{\pi}}^{i_{1: n}}\left(\boldsymbol{o}, \boldsymbol{a}^{i_{1: n}}\right)=\sum_{m=1}^{n} A_{\boldsymbol{\pi}}^{i_{m}}\left(\boldsymbol{o}, \boldsymbol{a}^{i_{1: m-1}}, a^{i_{m}}\right)
\end{equation}

This theorem guarantees that the joint action $a^{i_{1: n}}$ provides a positive advantage. To implement the this idea and address multi-agent collaborative tasks based on Dec-POMDP sequence modeling, we propose a collaborative algorithm for satellites swarm. This algorithm incorporates hybrid expert advice on top of the Proximal Policy Optimization algorithm.

The ``sequence to sequence" mechanism in constructing a multi-agent sequential decision-making paradigm significantly improves sample efficiency and addresses Dec-POMDPs-related challenges to some extent. Additionally, it is noteworthy that sequence models endow networks with long-term memory capabilities. In contrast to traditional MDP, the selection of actions now comprehensively considers previous historical state information rather than solely relying on the current state. This doesn't violate the Markov property, as the state within a certain time window remains unique. Furthermore, in scenarios with varying numbers and types of agents, sequence models showcase the ability to incorporate them into a unified solution by flexibly modeling the length of sequences, rather than treating different agent quantities as separate tasks. Simultaneously, this approach reduces the cumulative search complexity.

\subsection{Multi-agent Transformer}

Figure \ref{fig4} shows the designed Encoder-Decoder structure, where the parameters of the Encoder section, denoted by $\bm{\phi}$, encompass attention mechanisms and MLP. Additionally, residual connections are employed to prevent gradient vanishing and network degradation. Inputting observations from each agent yields a joint representation of these observations: $\left(\widehat{\boldsymbol{o}}^{i_{1}}, \ldots, \widehat{\boldsymbol{o}}^{i_{n}}\right)$. This encodes information from agents $\left(i_{1}, \ldots, i_{n}\right)$ over $\mathrm{T}$ preceding time steps, capturing advanced interactional dynamics among the agents. During the training phase, the encoder is used to approximate the value function and learn the embedded representation. The objective is to minimize the empirical Bellman error as follows:
\begin{equation}\label{eq10}
\begin{aligned}
&L_{\text {Encoder }}(\phi)\\
=&\frac{1}{T n} \sum_{m=1}^{n} \sum_{t=0}^{T-1}\!\left[\!R\left(\mathbf{o}_{t}, \mathbf{a}_{t}\right)\!+\!\gamma V_{\bar{\phi}}\left(\widehat{\mathbf{o}}_{t\!+\!1}^{i_{m}}\right)\!-\!V_{\phi}\left(\widehat{\mathbf{o}}_{t}^{i_{m}}\right)\!\right]\!^{2}
\end{aligned}
\end{equation}
where, $\bar{\phi}$ represents the parameters of the target network, which are updated periodically at fixed intervals.

The observation representation entering the decoder parameters is denoted by ${\boldsymbol{\theta}}$, and it shares a similar structure with the encoder network. This structure embeds the joint actions ${a^{i_{0: m-1}}}$ into a sequence for the decoding blocks, where ${m=\{1, \ldots n\}}$. The second attention module calculates attention between action heads and observation representations. The output from the decoding blocks is the sequence of joint action representations ${\left\{\widehat{\boldsymbol{a}}^{i_{0}: i-1}\right\}_{i=1}^{m}}$. Obtaining the policy ${\pi_{\theta}^{i_{m}}\left(\mathbf{a}^{i_{m}} | \widehat{\mathbf{o}}^{i_{1: n}}, \mathbf{a}^{i_{1: m-1}}\right)}$. To train the decoder, we minimize the following clipping PPO objective:
\begin{equation}\label{eq11}
\begin{aligned}
	&L_{\text {Decoder}}(\theta) \\
	\!=\!&\!-\!\frac{1}{{Tn}} \!\sum_{m\!=\!1}^{n}\! \!\sum_{t\!=\!0}^{T\!-\!1}\! \min \!\left(\!{r}_{t}^{i_{m}}(\theta) \hat{A}_{t}, \operatorname{clip}\!\left(\!{r}_{t}^{i_{m}}(\theta), 1 \!\pm\! \epsilon\!\right)\! \hat{A}_{t}\!\right)\!,\\
	&{r}_{t}^{i_{m}}(\theta)  =\frac{\pi_{\theta}^{i_{m}}\left(\mathrm{a}_{t}^{i_{m}} \mid \widehat{\mathbf{o}}_{t}^{i_{1: n}}, \hat{\mathbf{a}}_{t}^{i_{1: m-1}}\right)}{\pi_{\theta_{\mathrm{old}}}^{i_{m}}\left(\mathrm{a}_{t}^{i_{m}} \mid \widehat{\mathbf{o}}_{t}^{i_{1: n}}, \hat{\mathbf{a}}_{t}^{i_{1: m-1}}\right)}
\end{aligned}
\end{equation}
where, ${\hat{A}_{t}}$ represents the estimate of the joint advantage function. Generalized Advantage Estimation (GAE) \cite{bib34} is employed to obtain a robust estimator for the joint value function, denoted as ${\hat{V}_{t}=\frac{1}{n} \sum_{m=1}^{n} V\left(\hat{\mathrm{o}}_{t}^{i_{m}}\right)}$. It is noteworthy that the action generation processes during the inference and training phases differ. During inference, each action is generated autoregressively, starting from ${a^{i_{0}}}$ and ending with ${a^{i_{n-1}}}$, where ${a^{i_{m}}}$ is inserted into the decoder to generate ${a^{i_{m+1}}}$. In the training phase, actions ${a^{i_{1: n-1}}}$ are collected and stored in a replay buffer, and parallel computation is used to calculate the action ${a^{i_{1: n}}}$. The action decoding process by the decoder corresponds to the application of the multi-agent advantage decomposition theorem.

It is noteworthy that the agent ${i_{m}}$ conditions its new decision on the decisions of agents ${i_{1: m-1}}$ and optimizes a trust-region objective by adjusting its policy ratio Eq.(\ref{eq5}). Therefore, it monotonically increases the joint reward, similar to the sequential update scheme following the RNN architecture in HAPPO  \cite{bib15}. This scheme requires arranging the order of updates at each iteration to ensure that restricting the joint policy does not cause any agent to change its strategy (Nash equilibrium), inheriting the monotonic improvement guarantee of the PPO algorithm. However, unlike the HAPPO, the MAT model does not require $i_m$ to wait for the former to update, nor does it need to use the updated action distribution for importance sampling calculations. It parallelly computes their clipping objectives during the training phase, resulting in a lower time complexity compared to HAPPO.

\subsection{Expert mixture Transformer algorithm for collaborative multi-task in satellites swarm}

The ability to multitask effectively depends on the objectives and state space, as well as the relationship between expert networks and task representations. To address this issue, expert contributions are integrated through an attention mechanism. Instead of representing each task with a one-hot vector, we use a set of learnable independent tasks for embedding, which facilitates the extraction and fusion of expert knowledge. The expert network's attention weights indicate the importance of each expert's output for a given task. The soft mixture of expert network is defined as:
\begin{equation}\label{eq12}
h=\sum_{j=1}^{M} \alpha_{j} v_{j}=\sum_{j=1}^{M} \alpha_{j} V_{j} e_{j}
\end{equation}
Where ${v_{j}=V_{j} e_{j}}$ represents the transformation of expert output ${e_{j}}$ into expert value function ${v_{j}}$. To calculate attention weights, we use the softmax function to scale the dot product of the task query and expert keys as:
\begin{equation}\label{eq13}
\alpha_{j}=\frac{\exp \left(k_{j}^{T} q^{i}\right)}{\sum_{j=1}^{M} \exp \left(k_{j}^{T} q^{i}\right)}=\frac{\exp \left(e_{j}^{T} W_{j}^{T} q^{i}\right)}{\sum_{j=1}^{M} \exp \left(e_{j}^{T} W_{j}^{T} q^{i}\right)}
\end{equation}
Where ${k_{j}=W_{j} e_{j}}$ represents the transformation of expert output into expert key, and the task query ${q^{i}}$ is a set of independently trainable parameters unrelated to other task queries. Subsequently, the soft-combined expert networks are integrated into the backbone network to obtain the value function and actions. Specifically, as shown in Figure \ref{fig4}, there are two expert networks designed for extracting task-specific pursuit-attachment representations.

Our objective is to learn a general strategy for the pursuit-attachment two-stage task. To achieve this, expert networks is used to provide comprehensive guidance based on their experiences. However, due to varying granularity and difficulty levels across tasks, the convergence speed of training will differ, and simpler tasks may dominate the training process. To address this issue and train a well-balanced expert mixture network, we minimize the log-likelihood loss function of the experts:
\begin{equation}\label{eq14}
L_{ {reg }}(\varphi)=-\frac{1}{M} \sum_{j=1}^{M} \eta \log \left(\alpha_{j}+\varepsilon\right)
\end{equation}
where ${\eta}$ is a normalizing constant adjusting the regularization magnitude, and ${\varepsilon}$ is a small constant to prevent infinite values. We incorporate ${L_{r e g}}$ into the training process.

Furthermore, the proposed architecture utilizes soft combination that can enable the fusion process to be differentiable. And it allows for the expert networks and attention modules to be trained end-to-end. The proposed MAPPO-EMP for pursuit-attachment tasks of  satellites swarm can be presented as Algorithma 1:

\begin{algorithm}
	\caption{Expert mixture Transformer algorithm for collaborative multi-tasking in  satellites swarm}\label{alg1}
\begin{algorithmic}[1]
	\State \textbf{Input:} Learning rate $\boldsymbol{\alpha}$, mini-batch size B, number of agents n, episodes K, maximum steps per episode $\boldsymbol{T}$.
	\State \textbf{Initialization:} Orthogonal initialization Encoder $\{\phi_0\}$, Decoder $\{\theta_0\}$, Expert network $\{\varphi_0\}$; experience replay buffer $\mathcal{B}$.
	\For{$k = 1 \ldots K$}
	\For{$t = 1 \ldots T$}
	\State Collect sequences observation $[o_t^{i_1}, \ldots, o_t^{i_n}]$ in the environment.
        \Comment{The data collection phase}
	\State Encoder is fed with historical observation sequences $[o_t^{i_1}, \ldots, o_t^{i_n}], [o_{t-1}^{i_1}, \ldots, o_{t-1}^{i_n}], \ldots$ to generate representation sequences $[\hat{o}_t^{i_1}, \ldots, \hat{o}_t^{i_n}]$.
	\State Soft mixture of expert network knowledge to generating expert representation $h$
	\State Input $h$ to \textbf{Decoder}
	\For{$m = 0, 1, \ldots, n - 1$}
	\State According to the $a_t^{(i_m)} = \pi_{\theta}^{(i_m)} (\hat{o}_t^{i_{1:n}}, \hat{o}_{t-1}^{i_{1:n}}, \boldsymbol{a}^{i_{1:m-1}})$, where the input actions $a_t^{i_0}, \ldots, a_t^{i_{m-1}}$, the self-regressive decoder is used to infer the output $a_t^{i_m}$. This implicitly includes the current policy network $\pi_{\theta}^{(i_m)}$ of agent ($m$).
	\EndFor
	\State Execute the joint action $\boldsymbol{a}_t$ in the environment to obtain reward $R_t(\boldsymbol{o}_t, \boldsymbol{a}_t)$ and the next state and observation $s_{t+1}, \boldsymbol{o}_{t+1}$.
	\State Insert the tuple $(\boldsymbol{o}_t, \boldsymbol{a}_t, R(\boldsymbol{o}_t, \boldsymbol{a}_t), \boldsymbol{o}_{t+1})$ into experience replay buffer $\mathcal{B}$.
	\If {Complete task $i$ objectives and terminate environment}
	\State \textbf{break}
	\EndIf
	\EndFor
	\Comment{Training phase}
	\State Randomly sample batches of experiences of length $L$ from the experience replay buffer $\mathcal{B}$ with a batch size of B.
	\State Through the output layer of the encoder generate $V_\phi(\hat{o}^{i_1}), \ldots, V_\phi(\hat{o}^{i_n})$
	\State Calculate the encoder loss function $L_{\text{Encoder}}(\phi)$ using \textbf{Eq. \ref{eq10}}
	\State Using the GAE algorithm, calculate the joint advantage function $\hat{A}$ based on $V_\phi(\hat{o}^{i_1}), \ldots, V_\phi(\hat{o}^{i_n})$
	\State Sequentially generate their respective policies $[\pi_\theta^{i_1}, \ldots, \pi_\theta^{i_n}]$ for $\hat{o}^{i_1}, \ldots, \hat{o}^{i_n}$ and $a^{i_0}, \ldots, a^{i_{n-1}}$ through the decoder.
	\State Calculate the decoder loss function $L_{\text{Decoder}}(\theta)$ using \textbf{Eq. \ref{eq11}}
	\State Calculate the loss function $L_{\text{reg}}(\varphi)$ for the mixed expert network using \textbf{Eq. \ref{eq14}}
	\State Minimize $[L_{\text{Encoder}}(\varphi) + L_{\text{Decoder}}(\theta) + L_{\text{reg}}(\varphi)]$ using the gradient descent algorithm to update the parameters of the encoder network $\theta$, decoder network $\phi$, and mixed expert network $\varphi$.
	\EndFor
	\State \textbf{Output:} Optimal policy network $\pi_{\phi,\theta,\varphi}^*$
\end{algorithmic}
\end{algorithm}

It is noteworthy that the sequential decision-making process is independent of the decision order, allowing each agent to execute its decision independently. Afterward, agents receive rewards and new observations, which are stored as  experience tuples in the experience pool. The training phase follows. The globally centralized value network is only used during training; once training is complete, participants execute their respective action networks in a distributed manner. This architecture accommodates modeling of intelligent systems with a variable number and type of agents, thus establishing a multi-agent system with dynamic changes in the quantity of agents and mixed expert knowledge.

\section{NUMERICAL EXAMPLES AND RESULTS}\label{sec4}

\subsection{The process of experiments}

This section  presents an assessment of the proposed MAPPO-EMT method in a simulated orbital environment. A cooperative simulation environment for the satellites swarm are constructed by using PyTorch, leveraging the actual states of satellites. Specific environment parameters are outlined in Table \ref{tab2}. Subsequently, we provide details on the training and implementation of the proposed algorithm. The evasion strategies and learning capabilities for pursuit-attachment tasks are discussed within different scenarios. To highlight the features of our proposed method, MAPPO and MADDPG are employed in the same simulation environment. Finally, the obtained results are evaluated and discussed.

\begin{table}[ht]
	\centering
		\caption{Experience Setting}\label{tab2}
	\begin{tabular}{ccc}
		\hline
		\textbf{Parameter} & \textbf{Description} & \textbf{Value} \\
		\hline
		$a_0$ & reference orbit radius (km) & 42164 \\
        $\omega$& orbital angular velocity($\mathrm{rad} / \mathrm{s}$) & $7.27 \times 10^{-5} $ \\
		$l_r$ & side length (km) & 500 \\
		$r_e$ & evader radius (m) & 1\\
        $r_p$ & pursuer radius (m) & 1.5\\
        $r_{p, e}$ & attachment task radius (m) & 50\\
		$v_{\text{max}}$ & evader/pursuer max. speed (km/s)& 10 \\
        $\varpi_{\text {min }}$& minimum tolerance distance(m) & 0.1 \\
        $T_0$ & task time window (s) & $10^3$\\
		$\Delta t$ & time step (s) & 1 \\
		\hline
	\end{tabular}
\end{table}

The experiments are conducted on Windows 10 operating system powered by an Intel Core i7-13700K CPU @ 3.40 GHz, with training executed on an NVIDIA GeForce 3090 GPU. The orbital dynamic environment is developed using Python. The networks in the algorithm were constructed using PyTorch. To achieve improved convergence performance, we employ the built-in RMSProp optimizer for network training.

In the simulation environment, all the agents' positions are randomly initialized within a virtual boundary. They are then jointly driven by control forces and environmental factors, following the earlier established orbital dynamics rules as Eqs.\ref{eq3}-\ref{eq4} described.  Furthermore, all agents are subject to predefined constraints on maximum velocity and acceleration. It should be noted that the maximum acceleration limit for evaders is set at 1.2 times that of pursuers. This suggests that evaders have greater maneuverability and are therefore classified as 'advanced evaders.

During training, hyperparameters play a crucial role. For example, the outage penalty coefficient $\lambda$ helps to prompt the agent to find effective decision-making faster; the learning rate $\alpha$ is used to control the rate of model parameter updates, where a larger learning rate may lead to model oscillation, while a smaller one may slow down convergence; the discount factor $\gamma$ controls the influence of long-term rewards on current actions, with a higher discount factor implying a greater emphasis on long-term rewards. Careful adjustment of these hyperparameters can enable reinforcement learning models to learn better strategies, thereby improving training effectiveness.  More details are shown as Table \ref{tab3}.

\begin{table}[ht]
	\centering
	\caption{Hyperparameters Setting}\label{tab3}
	\begin{tabular}{ccc}
		\hline
		\textbf{Parameter} & \textbf{Description} & \textbf{Value} \\
		\hline
		$\lambda$ & outage penalty coefficient & 0.01 \\
		$\alpha$ & learning rate & $10^{-4}$ \\
		$\gamma$ & discount factor & 0.99 \\
		$\eta$ & soft update factor & 0.001 \\
		$e$ & batch size & 64 samples \\
		$\tau$ & target update rate & 0.002 \\
		$\mathcal{B}$ & Replay buffer capacity & $5 \times 10^4$ \\
		$\text{gae}\_\lambda$ & gae lambda parameter & 0.95 \\
		$\epsilon$ & Clipping strength & 0.05 \\
		\hline
	\end{tabular}
\end{table}

As described in section \ref{sec3}, the policy network parameters are shared among pursuers, making them homogeneous and sharing the same network. Therefore, the actor and critic networks are chosen with two hidden layers and ReLU activation functions to improve fitting performance. We utilize running mean and standard deviation to normalize rewards, and layer normalization is applied to inputs of networks. The network parameters are initialized with orthogonal weights and biases set to 0. The gain for the last action layer is set to 0.01. In addition, a RMSProp optimizer, a gradient-based optimization algorithm with adaptive learning rate characteristics, is used for accelerating the convergence process. It is also effective in handling sparse gradients.

 In this part, we describe a pursuit-attachment task with three satellites and one non-cooperative object.

\begin{figure}[!h]
	\centerline{\includegraphics[width=21.5pc]{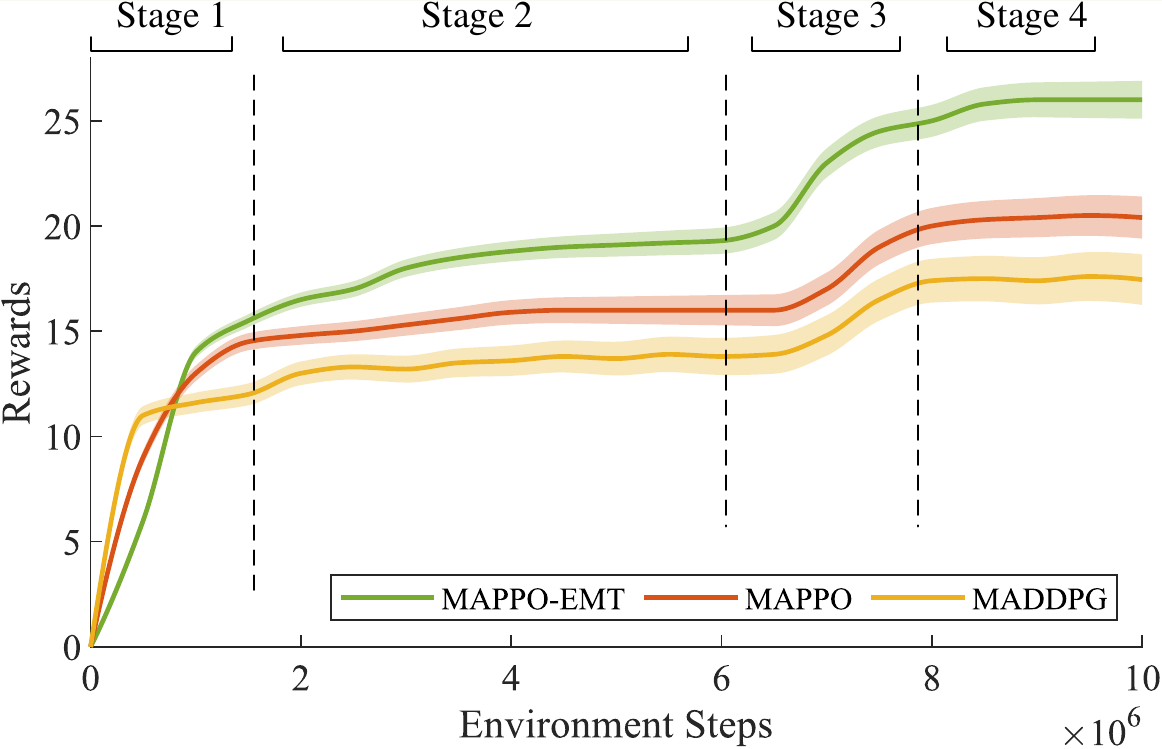}}
	\caption{Accumulated rewards comparison versus iterative number.}\label{fig5}
\end{figure}

The proposed algorithm performs excellently in terms of convergence and stability. As shown in Figure \ref{fig5}, it is clear that the rewards can be divided into four stages, during the initial training stage of the pursuit task (Stage 1), the rewards for all algorithms are relatively low due to the strong randomness in action selection. As the iteration steps increase, agents transition from the exploration stage to the Stage 2, it can be referred as pursuit task learning stage. The algorithm gradually converges, and the rewards stabilize incrementally. Similarly, Stages 3 and 4 represent the initial training and learning convergence stages of the attachment task, respectively. The two stages exhibits similar trends. In Stage 3, we utilize the expert-guided network in the MET structure to achieve sequential training and policy switching for the two-stage tasks. The algorithm's efficiency has been significantly enhanced.

Furthermore, it is evident that during the later stages of training, MADDPG's rewards still fluctuate due to the fact that MADDPG outputs a deterministic policy for action, while PPO outputs a policy or probability distribution. Notably, the proposed MAPPO-MET algorithm achieves stability after approximately 3 million iterations, outperforming other comparable algorithms in cumulative rewards during convergence. The MAPPO-MET algorithm utilizes the MET framework to efficiently incorporate historical temporal information and learn information transfer representations among agents. Therefore, it can better accomplish pursuit-attachment tasks compared to other algorithms.

\begin{figure}[!h]
	\centerline{\includegraphics[width=25pc]{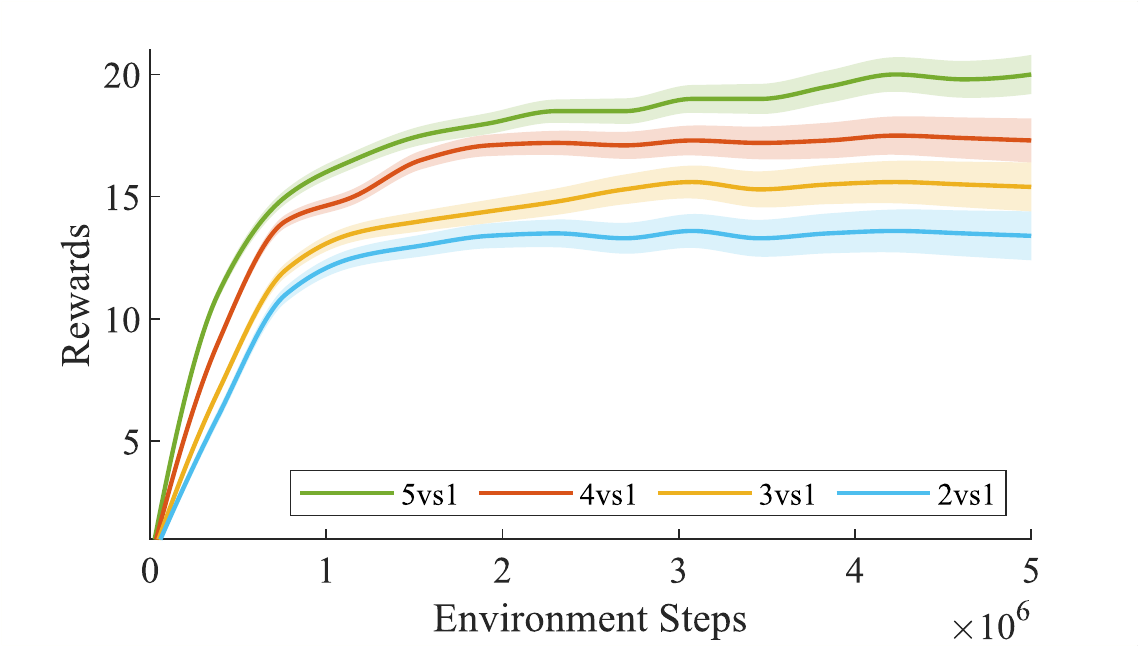}}
	\caption{Simulation experiment on the scalability of our proposed algorithm}\label{fig6}
\end{figure}

To explain how the algorithm addresses scalability issues caused by changes in the number of agents, we conducted a series of simulation experiments. Figure \ref{fig6} shows that our algorithm can be trained and deployed flexibly without changing the network structure when the number of agents ranges from 2 to 5, demonstrating its generality and flexibility. Subfigures \ref{fig9d} and \ref{fig9f} respectively respectively show that the pursuit tasks for '5vs1' and '4vs2' of the pursuers and evaders. It can be observed that when there are two pursuers, the average reward converges to approximately half of the reward achieved when there are three pursuers. This suggests that the collaborative capture capability reaches its maximum with two pursuers. When the number of additional pursuers exceeds three, the improvement in capture task effectiveness becomes less significant. This is due to the increased penalty for avoiding collisions between agents, which hinders the increase in rewards. Furthermore, deploying more pursuers results in increased computational resource consumption. It is important to note that even when deploying a network trained in a 3vs1 scenario to a 4vs1 environment, satisfactory results can still be achieved.

\begin{figure}[!h]
	\centerline{\includegraphics[width=19.5pc]{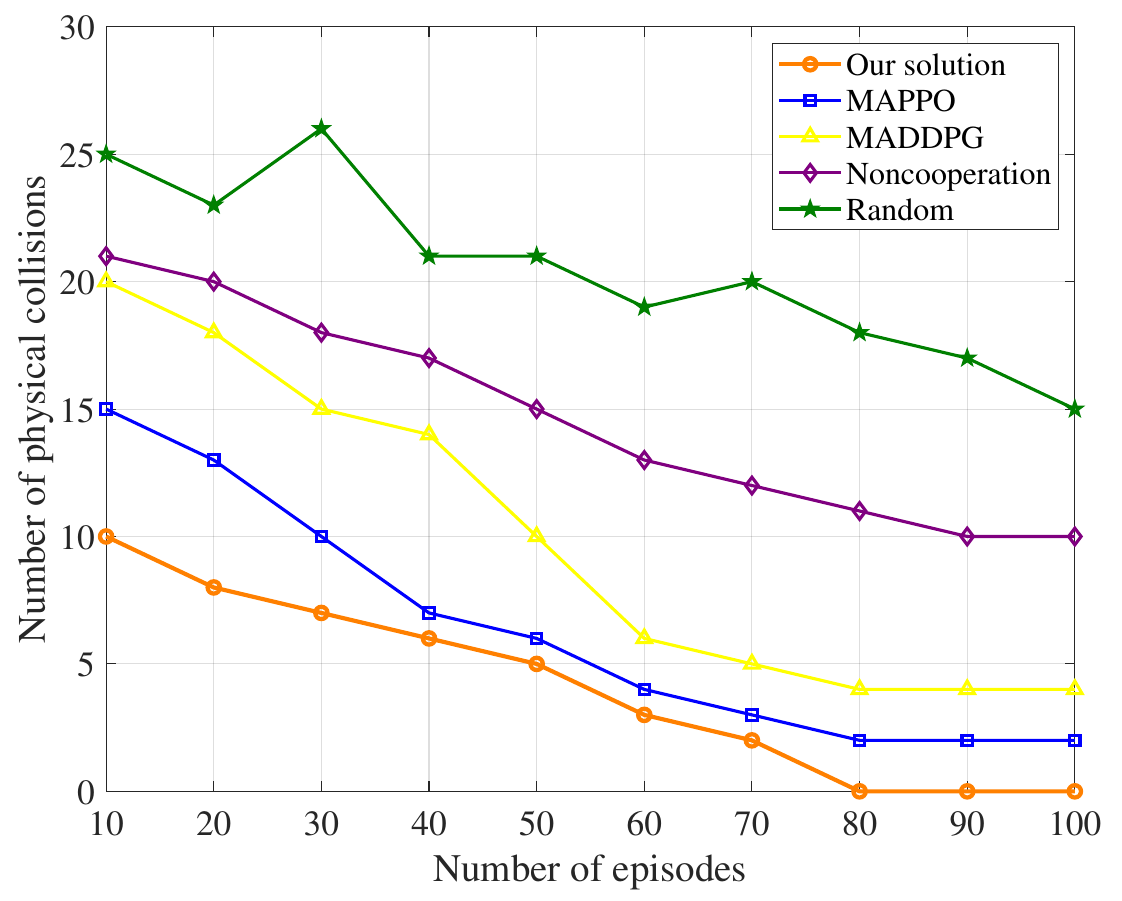}}
	\caption{Number of physical collisions versus number of episodes}\label{fig7}
\end{figure}

In Figure \ref{fig7}, the performance of each algorithm in collision avoidance is presented. The physical collisions can be defined based on the minimum tolerance distance between agents, represented as $\varpi_{{min}}$. And one can record the number of collisions during the process of training. We observed that all approaches demonstrated a certain degree of reduced collision frequency. Other methods failed to completely address the challenge of collision avoidance. In contrast, our proposed algorithm successfully achieved collision-free instances when the episode count reached approximately 80. This is because we have transformed safety hard constraints into learnable soft constraints in the reward design.

\begin{figure}[!h]
	\centering
	\begin{subfigure}[b]{0.5\textwidth}
		\centering
		\includegraphics[width=18pc]{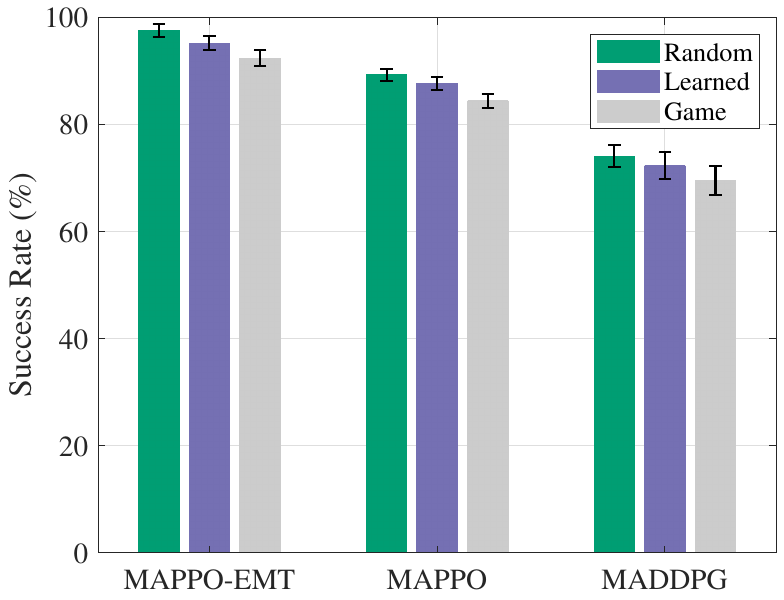}
		\caption{pursuit task}
		\label{fig8a}
	\end{subfigure}
	\begin{subfigure}[b]{0.5\textwidth}
		\centering
		\includegraphics[width=18pc]{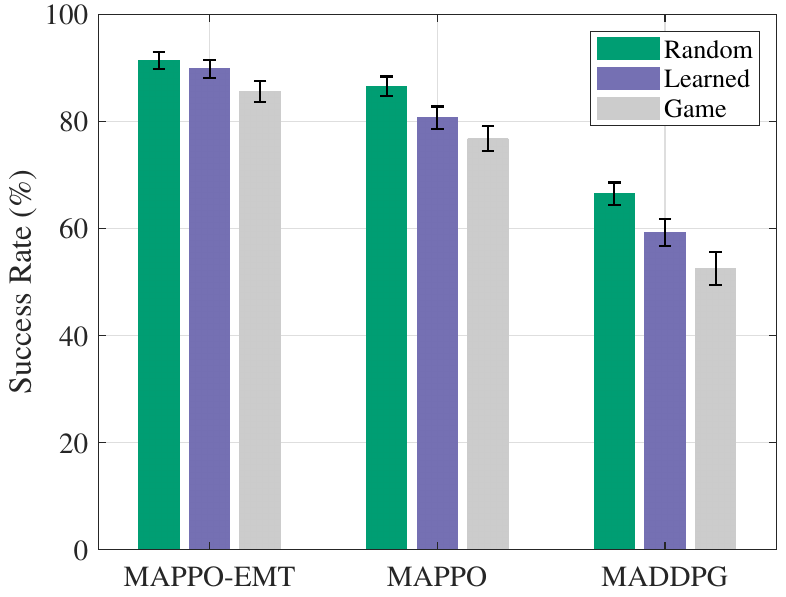}
		\caption{attachment task}
		\label{fig8b}
	\end{subfigure}
	\caption{Success rates of two tasks in three scenarios.}
	\label{fig8}
\end{figure}

In order to evaluate the deployment and execution efficiency of our solution, three benchmarks are used to compare performances:

1) Random evasion strategy: the evader randomly uses impulses to control his movement. 

2) Pre-trained avoidance strategy: We apply our algorithm to learn the avoidance strategy. During the training process, both the pursuer and the evader are controlled by the algorithm, and collisions between them are penalized. After pre-training, the evader will acquire a certain level of evasion strategy. 

3) Game adversarial strategy: Let the evader and pursuer enter a competitive game, learning both evasion and pursuit strategies simultaneously through our algorithm.

Despite the need for approximately 4 hours and 8 million iterations for training, the post-training network no longer relies on the critic network. Deploying the trained policy network in distributed intelligent agents, we can achieve collaborative pursuit tasks with minimal computational overhead. In each decision step, a single forward pass through the policy network produces respective actions, significantly reducing computational expenses compared to traditional control algorithms. Subsequently, we conduct 10,000 Monte Carlo task success experiments within a limited time window in three pursuit-evasion scenarios: random evasion strategy, pre-trained evasion strategy, and adversarial game strategy. The results, as illustrated in Figure \ref{fig8}, show that all algorithms perform well in the random evasion scenario. However, under the pre-trained evasion strategy, MADDPG exhibits slightly suboptimal performance. In the adversarial game scenario, our proposed algorithm outperforms the alternatives.

\begin{figure}[!h]
	\centerline{\includegraphics[width=22pc]{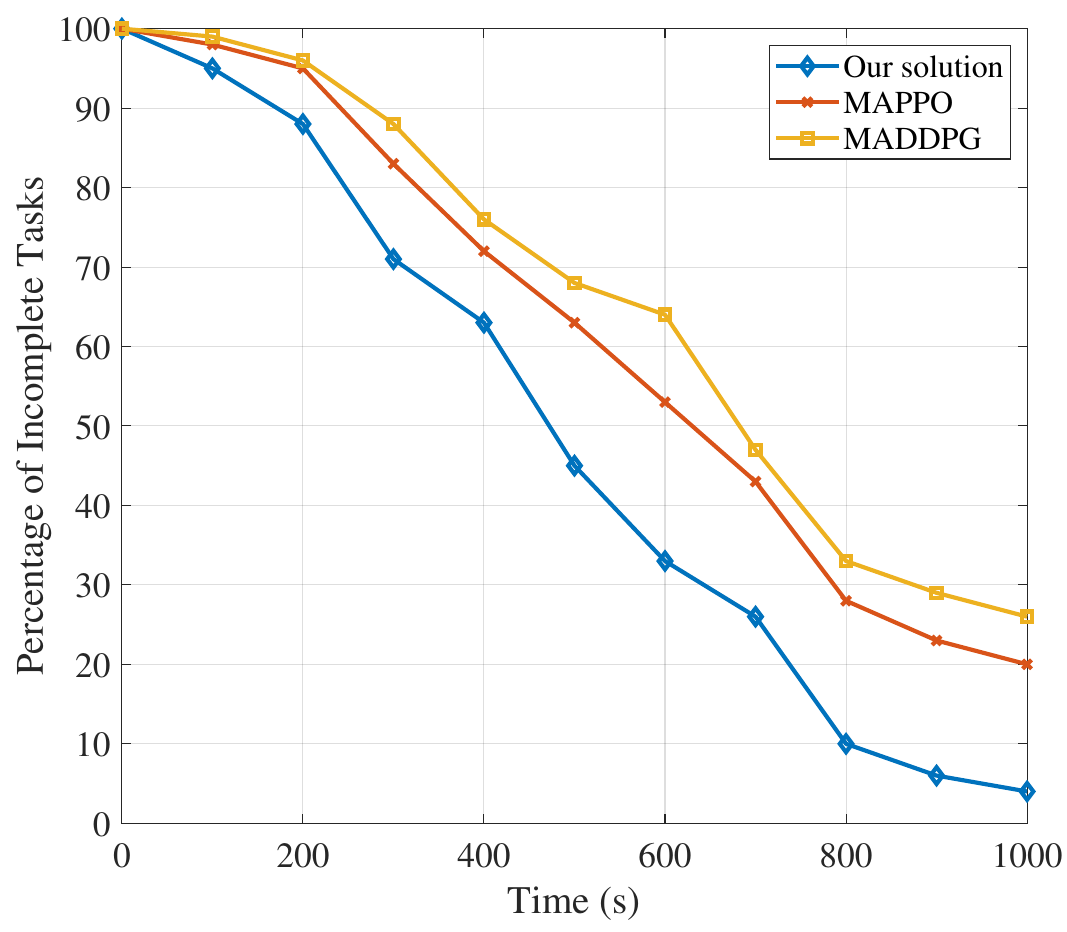}}
	\caption{Percentage of incomplete tasks versus time}\label{fig8.5}
\end{figure}

\begin{table}[ht]
	\centering
	\caption{Comparison of fuel consumption }\label{tab4}
        \begin{tabular}{|c|c|c|c|c|c|}
        \hline
        satellite & $\Delta v_E$ & $\Delta v_{P_1}$ & $\Delta v_{P_2}$ & $\Delta v_{P_3}$ & $\Delta v_P$ \\
        benchmarks&$(\mathrm{m} / \mathrm{s})$ & $(\mathrm{m} / \mathrm{s})$ & $(\mathrm{m} / \mathrm{s})$ & $(\mathrm{m} / \mathrm{s})$ & $(\mathrm{m} / \mathrm{s})$ \\

        \hline
        random & 44.09 & 54.37 & 51.99 & 54.09 & 160.45 \\
        \hline
        pre-trained & 76.83 & 70.76 & 68.64 & 68.07 & 207.47 \\
        \hline
        game & 140.26 & 130.74 & 127.11 & 126.79 & 384.64 \\
        \hline
        \end{tabular}
\end{table}

In order to evaluate the efficiency of the deployment algorithms in executing the tasks, we compared the completion times of three algorithms for pursuit-attachment tasks in the context of a pre-trained evasion strategy. Due to the presence of randomness, the completion times of tasks vary. We conducted 1,000 Monte Carlo experiments, observing the remaining number of incomplete tasks at certain time intervals within the task time window. As illustrated in Figure \ref{fig8.5}, our method exhibited the most rapid decline within the designated task time window.
Furthermore, as the cumulative amount of speed change is directly proportional to the maneuvering fuel consumption, we use the cumulative amount of speed change to evaluate the maneuvering consumption performance. For example, Table \ref{tab4} shows the progressive attachment tasks under three benchmark settings, with the cumulative velocity changes of the subsatellites and target satellites and the total fuel consumption of the swarm.  It can be seen that the consumption of the three satellites within the swarm is relatively similar.  Among the three benchmarks, the game consumes the most fuel, approximately 184$\%$ more than the others.  This is due to the non-cooperative satellites having advanced evasion strategies, and the two sides engaging in intense competition.

\begin{figure*}[!h]
	\centering
	\begin{subfigure}[b]{0.45\textwidth}
		\centering
		\includegraphics[width=15pc]{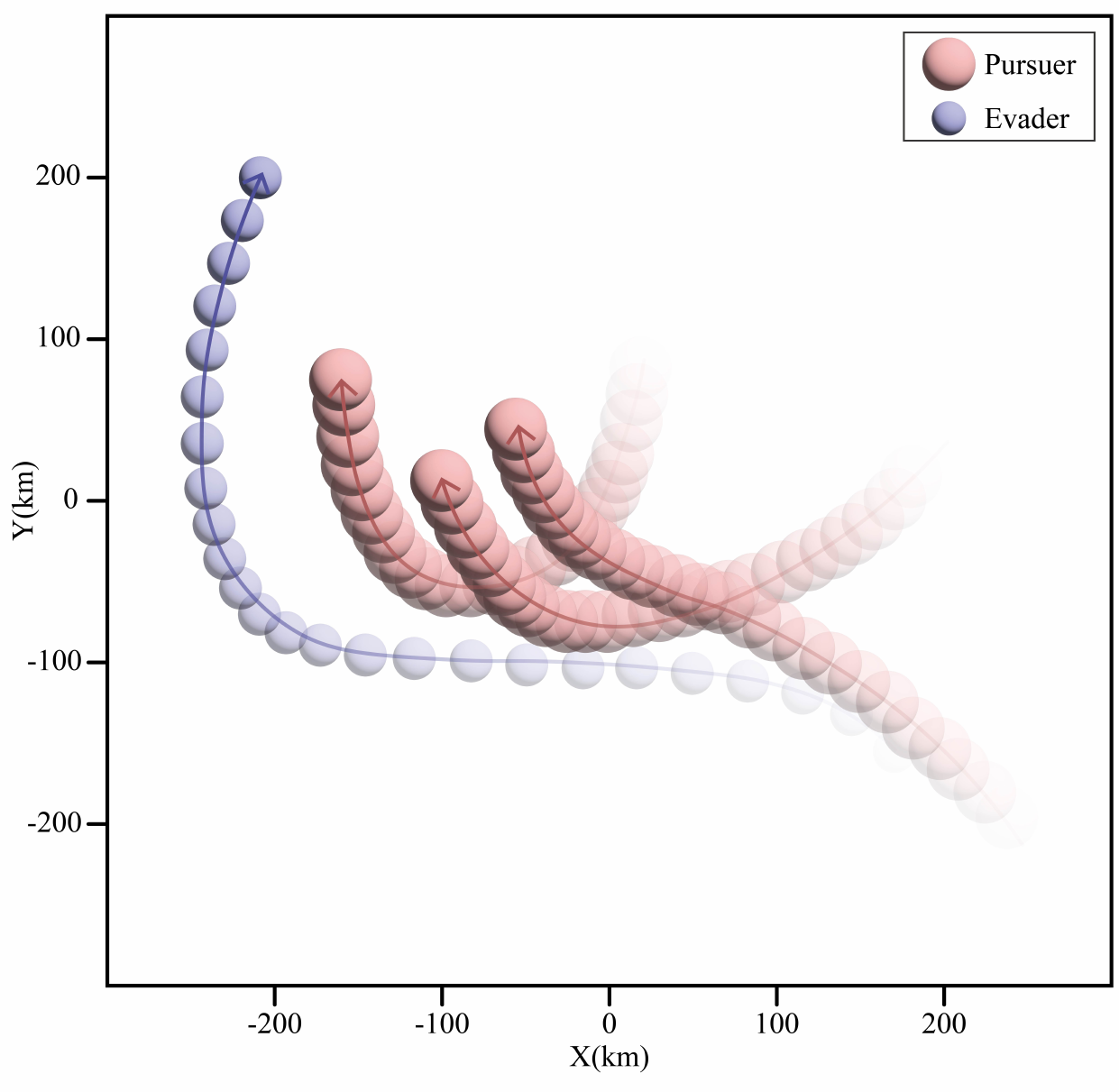}
		\caption{3 vs 1 pursuit tasks success}
		\label{fig9a}
	\end{subfigure}
	\begin{subfigure}[b]{0.45\textwidth}
		\centering
		\includegraphics[width=15pc]{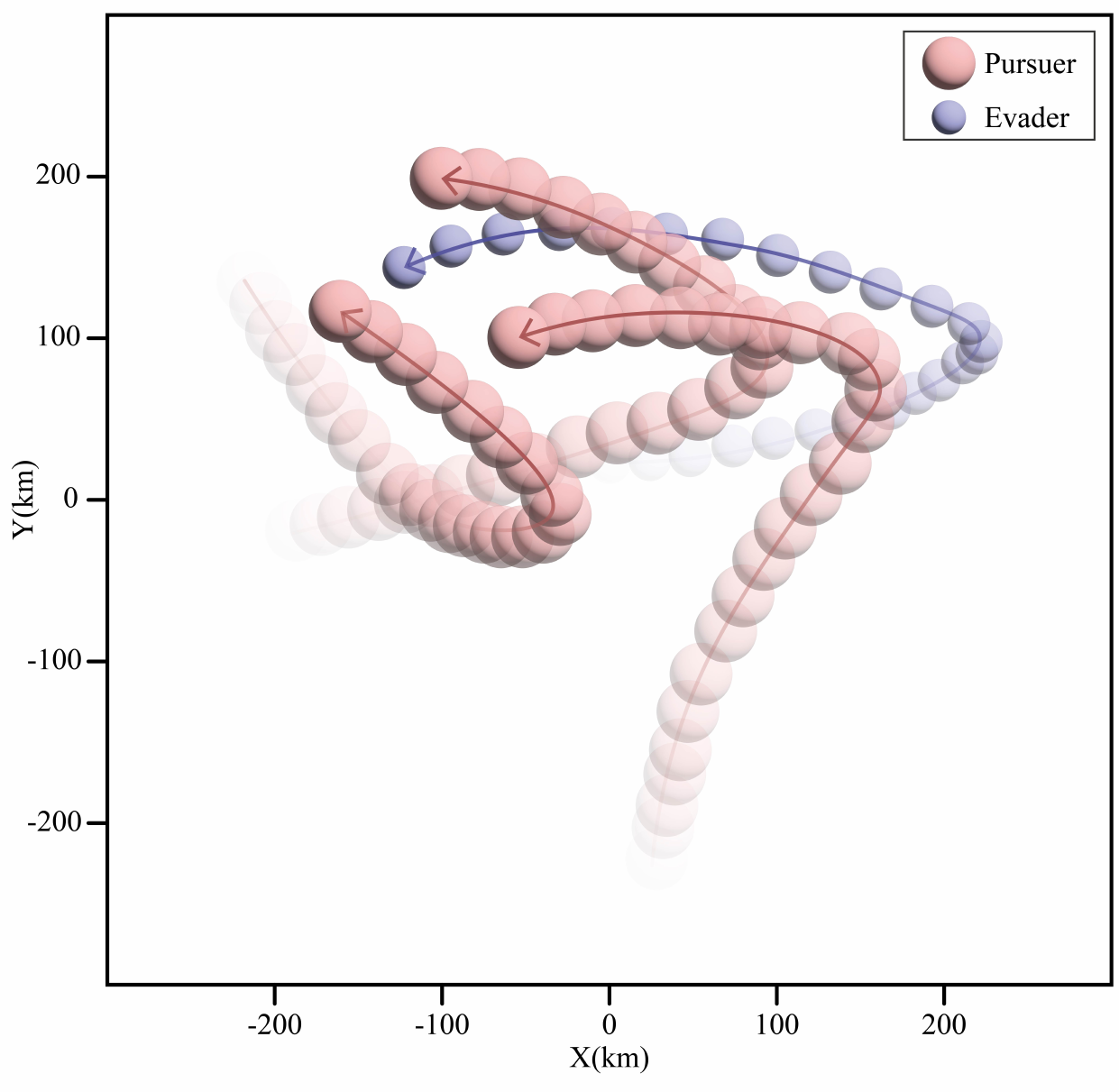}
		\caption{3 vs 1 pursuit tasks success}
		\label{fig9b}
	\end{subfigure}
		\begin{subfigure}[b]{0.45\textwidth}
		\centering
		\includegraphics[width=15pc]{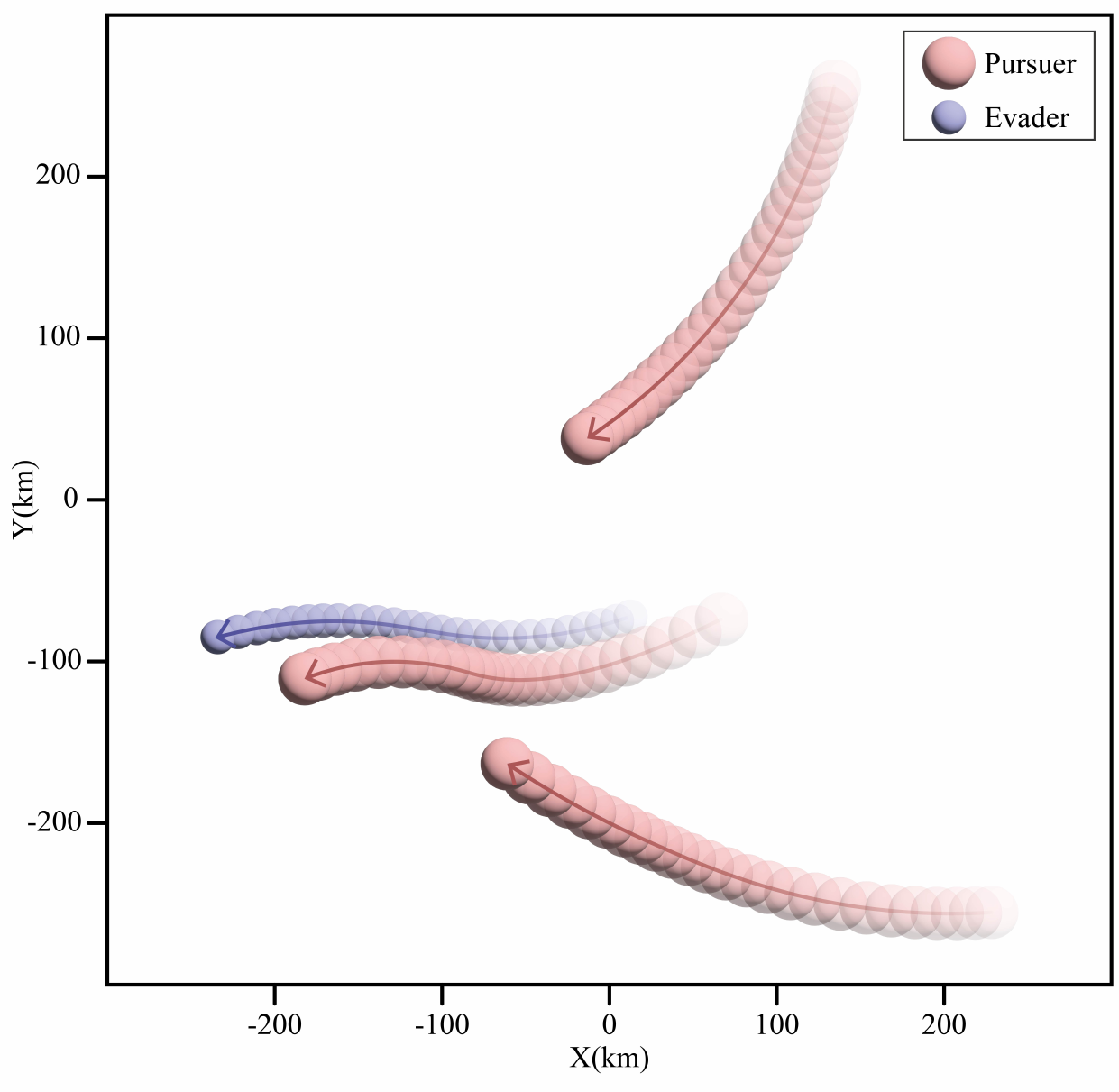}
		\caption{3 vs 1 pursuit tasks failed}
		\label{fig9c}
	\end{subfigure}
		\begin{subfigure}[b]{0.45\textwidth}
		\centering
		\includegraphics[width=15pc]{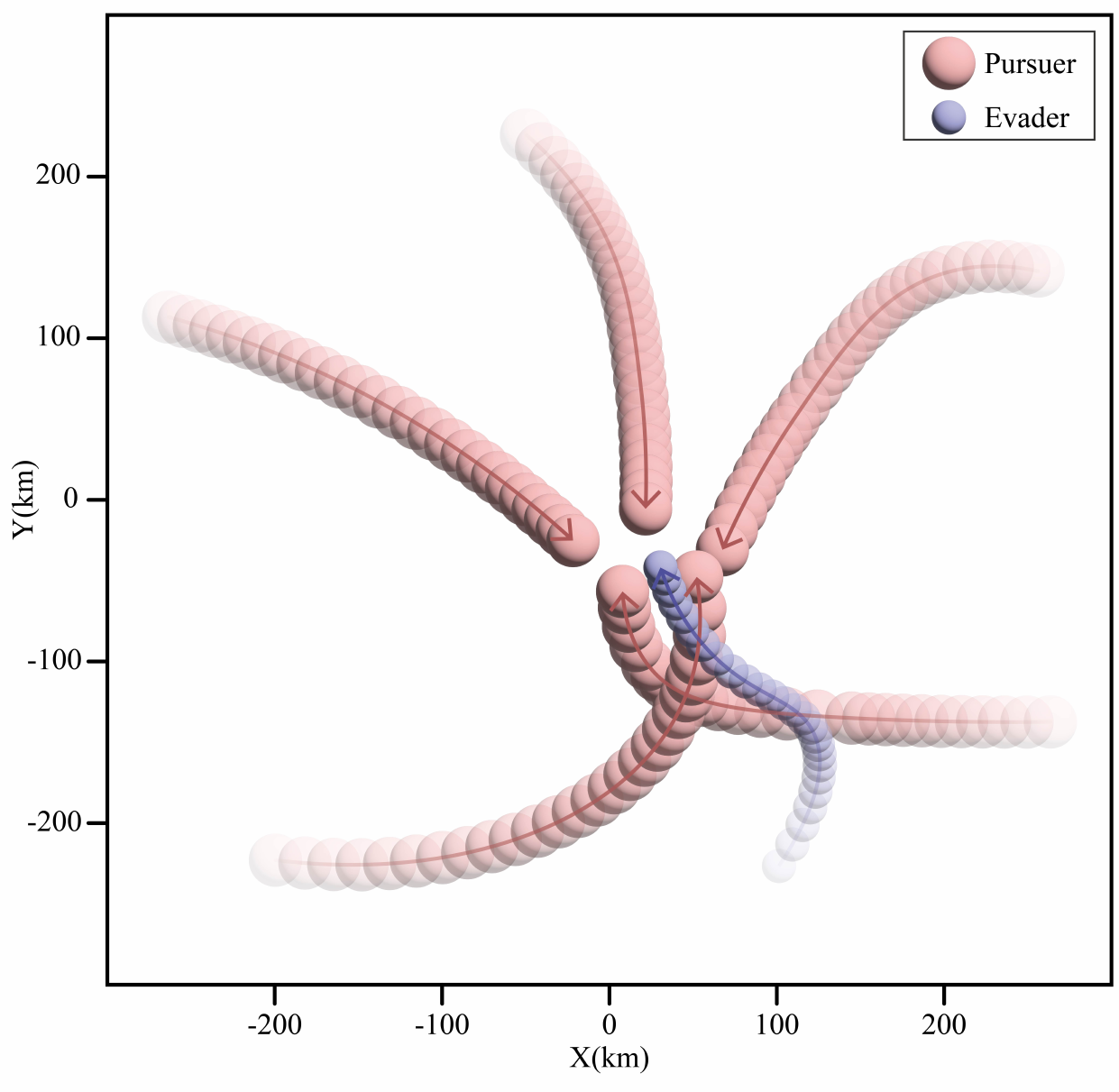}
		\caption{5 vs 1 pursuit tasks success}
		\label{fig9d}
	\end{subfigure}
	\begin{subfigure}[b]{0.45\textwidth}
		\centering
		\includegraphics[width=15pc]{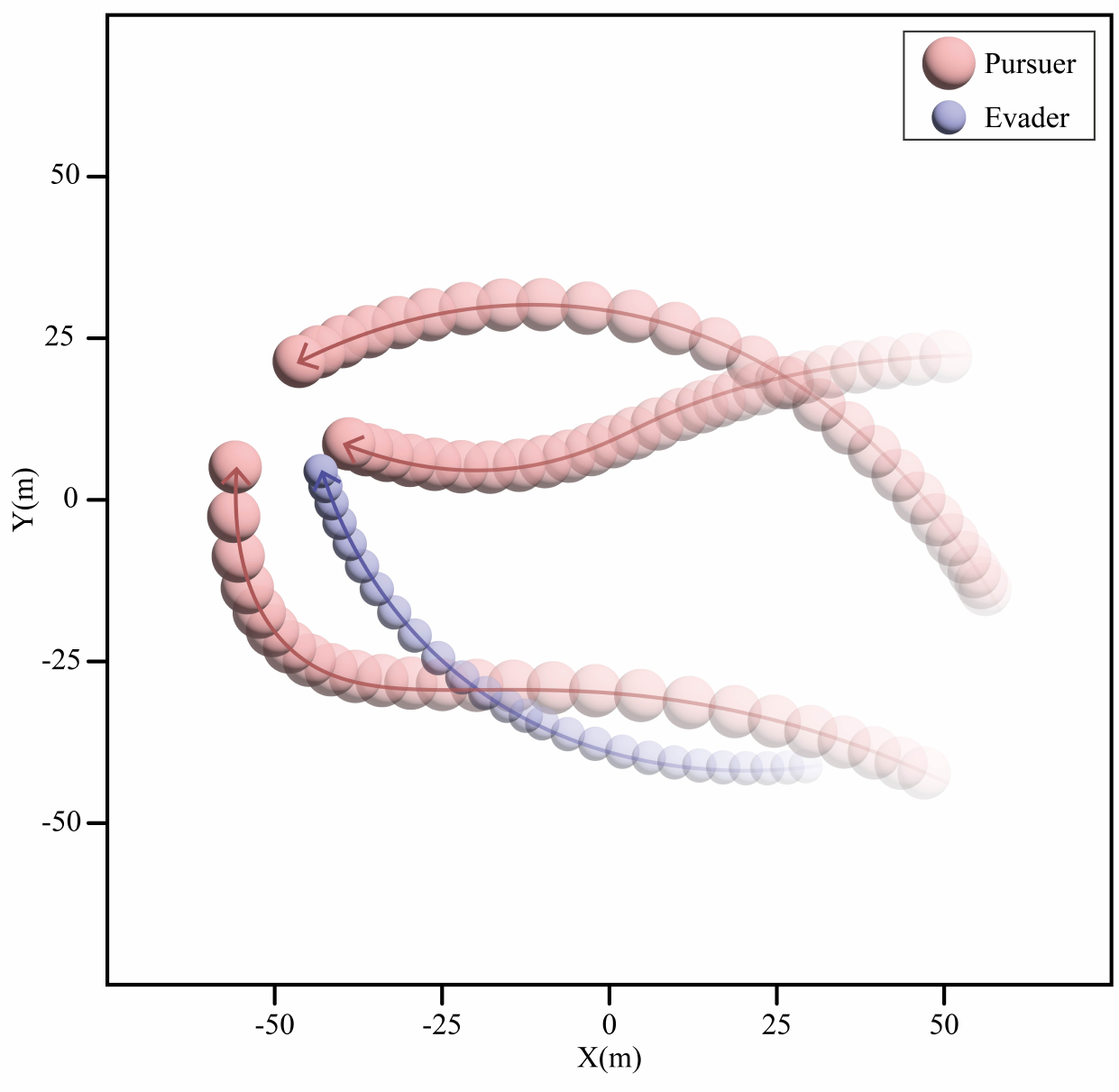}
		\caption{3 vs 1 attachment  tasks success}
		\label{fig9e}
	\end{subfigure}
	\begin{subfigure}[b]{0.45\textwidth}
		\centering
		\includegraphics[width=15pc]{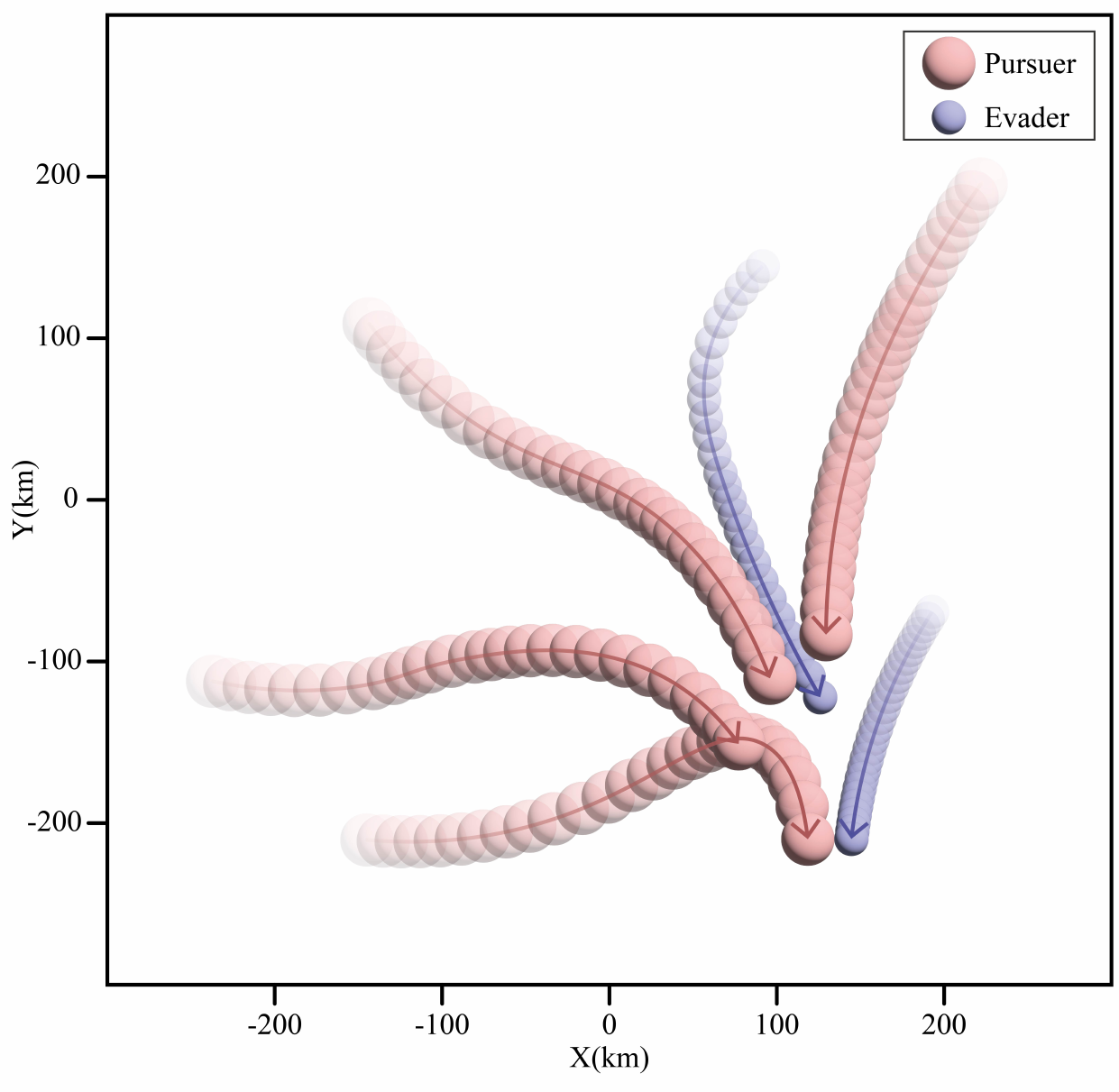}
		\caption{4 vs 2  pursuit tasks }
		\label{fig9f}
	\end{subfigure}
	\caption{Motion trajectory of pursuers and evaders in the simulation environment}
	\label{fig9}
\end{figure*}
\newpage

Figure \ref{fig9} displays two successful cases of pursuit tasks: subfigure \ref{fig9a} illustrates the scenario where the pursuer (our satellites swarm) drives the evader (Non-cooperative satellite) to the virtual boundary, and subfigure \ref{fig9b} shows that the pursuers surrounding the evader. Additionally, subfigure \ref{fig9c} illustrates a failure scenario in the pursuit task. subfigure \ref{fig9d} displays a schematic diagram of successful pursuit when there are five pursuers in the  satellites swarm, and the skill level of evaders is set to 1. This demonstrates that our sequence modeling Transformer architecture provides the algorithm with strong scalability and generalization. subfigure \ref{fig9e} illustrates the successful execution of attachment tasks, indicating the guidance provided by our Expert Network. As defined the Section \ref{sec3}, the MAPPO-EMT algorithm can share neural network parameters based on task similarity and utilize an expert network to achieve soft switching between two-stage task executions. Additionally, experiments are conducted with multiple evaders, using 4 pursuers and 2 evaders, as depicted in subfigure \ref{fig9f}, the algorithm's effectiveness are demonstrated furtherly.

For this simulation, three satellites conduct a pursuit-attachment operation on a non-cooperative satellite using MAPPO-EMT and predefined configurations. As shown in previously stated, MAPPO-EMT utilizes sequence modeling and an expert knowledge network, combined with the monotonic improvement guarantee of the inherited PPO algorithm. These approaches allow for the resolution of scalability issues that arise from variable numbers of observable entities, as well as the alleviation of convergence difficulties and instability as shown in figure \ref{fig5}-\ref{fig6}. Compared to MADDPG, MAPPO-EMT does not employ a clipping mechanism during policy updates. This lack of clipping can result in significant policy changes during update steps, making the algorithm more susceptible to noise and relatively unstable during training. Additionally, the proposed approach utilizes multiple trajectories for policy updates, which improves the algorithm's sample efficiency by allowing for a more comprehensive utilization of sampled data. In contrast, MADDPG typically uses an offline experience replay mechanism, which may have limitations in terms of sampling efficiency.

\section{CONCLUSION}\label{sec5}
This manuscript investigates the collaborative problem of pursuit-attachment tasks in a satellites swarm. Firstly, we conducted modeling based on CW orbital kinematics and cooperative computation. Then, we designed a multi-agent reinforcement learning algorithm framework guided by experts. 
Sequence modelling based on transformers increases the capacity of memory-augmented policy networks
Expert networks use similarities between tasks to guide task switching, improving sample efficiency when training multi-task policies.
The algorithm is capable of adapting to complex orbital kinematic environments and responding to dynamic satellites, enabling real-time continuous task execution. Compared with existing benchmarks, the results indicate that the proposed algorithm converges rapidly, possesses lower complexity, and exhibits algorithmic scalability. It can be flexibly deployed and executed according to the needs of pursuit-attachment tasks.

\appendix
\section{Appendix 1}

$\begin{aligned} & \widetilde{\boldsymbol{A}}=\left[\begin{array}{cccccc}0 & 1 & 0 & 0 & 0 & 0 \\ 3 \omega_0^2 & 0 & 0 & 2 \omega & 0 & 0 \\ 0 & 0 & 0 & 1 & 0 & 0 \\ 0 & -2 \omega & 0 & 0 & 0 & 0 \\ 0 & 0 & 0 & 0 & 0 & 1 \\ 0 & 0 & 0 & 0 & -\omega_0^2 & 0\end{array}\right] \\ & \widetilde{\boldsymbol{B}}=\left[\begin{array}{lll}0 & 0 & 0 \\ 1 & 0 & 0 \\ 0 & 0 & 0 \\ 0 & 1 & 0 \\ 0 & 0 & 0 \\ 0 & 0 & 1\end{array}\right] \\ & \end{aligned}$

\bibliographystyle{IEEEtaes}
\bibliography{ref}

\end{document}